\documentclass[prl,reprint,superscriptaddress,nobibnotes]{revtex4-1}

\usepackage[dvipdfmx]{graphicx}
\usepackage{amssymb, ascmac, amsthm,amsmath,latexsym}

\begin{document}
	\title{Topological origin of universal few-body clusters in Efimov physics}
	
	\author{Yusuke Horinouchi}
	\email[E-mail: ]{yusuke@cat.phys.s.u-tokyo.ac.jp}
	\affiliation{Department of Physics, University of Tokyo, Hongo 7-3-1, Bunkyo-ku, Tokyo 113-0033, Japan}
	\author{Masahito Ueda}
	\affiliation{Department of Physics, University of Tokyo, Hongo 7-3-1, Bunkyo-ku, Tokyo 113-0033, Japan}
	\affiliation{RIKEN Center of Emergent Matter Science (CEMS), Wako, Saitama 351-0198, Japan}
	
	\date{\today}
	\begin{abstract}
		Efimov physics is renowned for the self-similar spectrum featuring the universal ratio of one eigenenergy to its neighbor.  Even more esoteric is the numerically unveiled fact that every Efimov trimer is accompanied by a pair of tetramers. Here we demonstrate that this hierarchy of universal few-body clusters has a topological origin by identifying the numbers of universal 3- and 4-body bound states with the winding numbers of the renormalization-group limit cycle in theory space.  The finding suggests a topological phase transition in mass-imbalanced few-body systems which should be tested experimentally.
	\end{abstract}
	\maketitle
	Universality in physics often refers to a situation in which apparently distinct systems show the same low-energy behavior. A prominent example is the critical phenomena, where different physical systems are grouped into a set of universality classes sharing the same critical exponents. The modern foundation for understanding the universality is established by the renormalization group (RG) \cite{wilson1974}, which allows us to investigate a change of a system viewed at different distance scales by following an RG flow of system-parameters generated by a recursive coarse graining of the system. In particular, the universality classes of critical phenomena can be categorized by the fixed points of the RG flow, which represent the scale invariance of the second-order phase transition at which no characteristic length scale is present. \par
\begin{figure}[t]
  			\begin{center}
  			 	\includegraphics[width=250pt]{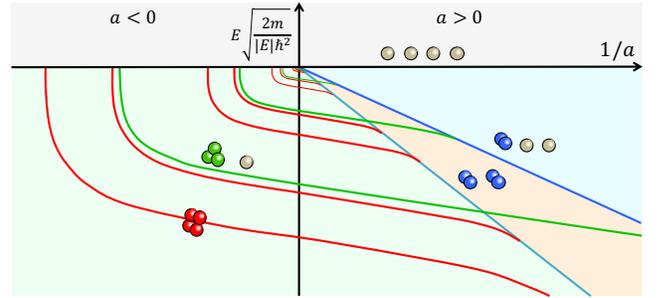}
				\caption{Schematic illustration of the energy spectrum of four identical bosons near the scattering resonance. The abscissa shows the inverse $s$-wave scattering length $a^{-1}$ and the ordinate shows the square root of the energy eigenvalues. The Efimov trimers (green-colored curves) are related to one another by a scaling factor of $22.7$, and at the scattering resonance $a=\pm\infty$, the energies $E_{4b}^{(1)}$ and $E_{4b}^{(2)}$ of two tetramers (red-colored curves) associated with a trimer can be related to the trimer energy $E_{3b}$ by $E_{4b}^{(1)}/E_{3b}\simeq 1.01$ and $E_{4b}^{(2)}/E_{3b}\simeq 4.58$, respectively \cite{stecher2009}. A more detailed energy spectrum can be found in Refs.~\cite{deltuva2011, deltuva2013a, deltuva2013b}.}\label{fig:fig1}
  			\end{center}
		\end{figure}

	Yet another characteristic RG flow can be found in universal quantum few-body physics that feature discrete scale invariance. Here we consider the Efimov effect \cite{efimov1970} and its 4-body extension \cite{platter2004, hammer2007, stecher2009} in which resonantly interacting 3 and 4 bosons form an infinite series of universal 3- and 4-body bound states that feature the self-similar spectrum: the 3-body bound states (Efimov trimers) are related to one another by a scaling factor of $(22.7)^2$, and each Efimov trimer is accompanied by two 4-body bound states (see Fig.~\ref{fig:fig1}). This discrete scale invariance makes the Efimov physics a prime example of the RG limit cycle \cite{wilson1971, glazek2002, bedaque1999, bedaque1999b, moroz2009}, where an RG flow forms a periodic circle rather than converges to a fixed point. 
	 Because of the universality and the uniqueness, much theoretical \cite{efimov1970, efimov1973, petrov2003, nishida2009, hammer2010, nishida2013, pal2013} and experimental \cite{kraemer2006, knoop2009, zaccanti2009, gross2009, pollack2009, huckans2009, barontini2009, pires2014, huang2014, tung2014, kunitski2015} efforts have been devoted to reveal the existence of the Efimov trimers in a rich variety of systems. Also, the existence of the 4-body companions associated with Efimov trimers has been confirmed both numerically \cite{platter2004, hammer2007, stecher2009, deltuva2010, deltuva2011, deltuva2013a, deltuva2013b} and experimentally \cite{ferlaino2009}.\par
	 Despite such extensive research, there remains an as yet unresolved fundamental question: How is the universal 4-body physics related to the RG limit cycle? In the following, we answer this question by showing that the hierarchical structure of the few-body clusters has a topological origin in terms of the RG limit cycle, as we conjectured previously \cite{horinouchi2015}. In particular, we numerically demonstrate that the one-to-two ratio between the number of Efimov trimers and that of the associated 4-body bound states can be understood in terms of topological numbers in theory space.\par
		\begin{figure*}[t]
  			\begin{center}
  			 	\includegraphics[width=480pt]{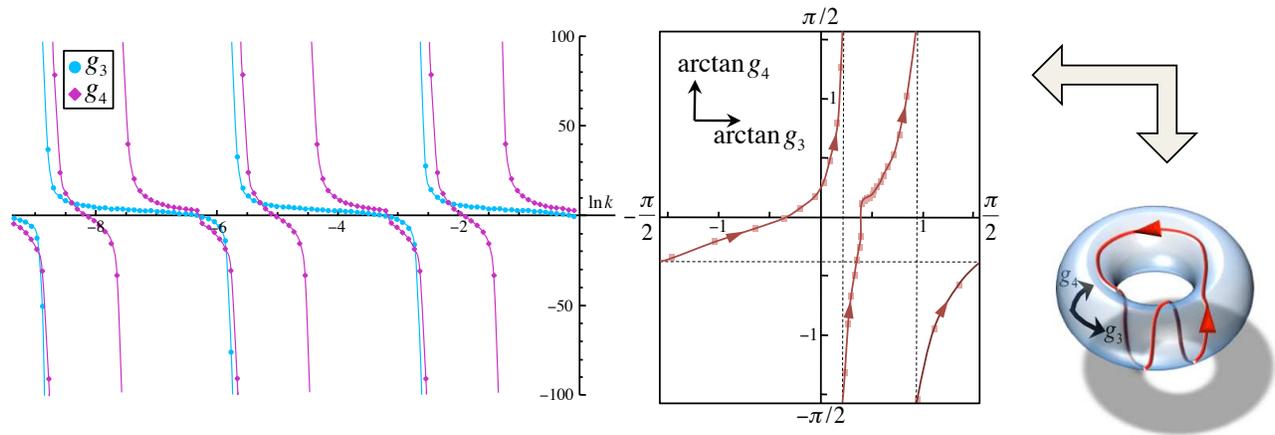}
				\caption{{\bf a}, Cutoff $k$ dependences of the 3-body and 4-body coupling constants $g_3$ and $g_4$ at the scattering resonance $1/a=0$. The abscissa shows the logarithm of the cutoff $k$. The blue dots and the purple diamonds show $g_3/2$ and $g_4/250$, where the multiplication factors $1/2$ and $1/250$ are introduced to display the two flows simultaneously. {\bf b}, RG flow in the theory space of the 3-body and 4-body coupling constants. The abscissa shows ${\rm arctan}(g_3/16)$ and the ordinate shows ${\rm arctan}(g_4/6250)$, where we use arctan to display diverging coupling constants, and $g_3$ and $g_4$ are multiplied by 1/16 and 1/6250 for the sake of simultaneous display. The brown curve shows the RG limit cycle obtained by eliminating the cutoff $k$ dependence of $g_3$ and $g_4$ from Fig.~\ref{fig:fig2}{\bf a}. If we glue the edges of the figure to form a torus, the limit cycle winds once in the $g_3$ direction and twice in the $g_4$ direction. Thus the numbers of 3-body and 4-body bound states correspond to the topological winding numbers of the limit cycle. {\bf c}, Schematic illustration of the topological nature of the limit cycle described in Fig.~\ref{fig:fig2}{\bf b}. The figure {\bf c} is adapted from Fig.~5 of Ref.~\cite{horinouchi2015} with proper modifications.}\label{fig:fig2}
  			\end{center}
		\end{figure*}
	\subsection{Topological winding numbers of the limit cycle}
		The renormalization group (RG) recursively performs a coarse graining transformation of an action, in accordance with the cutoff-parameter $k$ which characterizes the `resolution' of the coarse graining. 
		The effect of the RG transformation on the 3- and 4-body physics can be seen from the 3- and 4-body coupling constants $g_3$ and $g_4$, which are defined as $k$-dependent particle-dimer and particle-trimer scattering amplitudes, respectively. To obtain the RG flows of $g_3$ and $g_4$, we use the functional renormalization group (FRG) which enables us to deal with such strongly correlated situations as the Efimov physics.\par
		Previous FRG analyses \cite{moroz2012, avila2013, avila2015} on the resonantly interacting 4 bosons suffer spurious 4-body bound states due to the point-like approximation where the momentum dependence of Green's functions are almost disregarded. To avoid such an artifact, we take into account the full momentum dependence of the Green's functions, with the aid of the separable pole approximation \cite{alt1970, grassberger1967}, which enables one to represent a Green's function with a tractable function. \par
		 In Fig.~\ref{fig:fig2}{\bf a}, $g_3$ and $g_4$ at the scattering resonance are plotted as functions of the cutoff $k$, which demonstrates that $g_3$ and $g_4$ both have the log-periodic $k$ dependences to form an RG limit cycle. Furthermore, $g_4$ flows twice from $-\infty$ to $\infty$ every time $g_3$ flows from $-\infty$ to $\infty$, indicating that there appear two 4-body bound states associated with one Efimov trimer, since the divergence of $g_3$ and $g_4$ can be identified with the appearance of the 3- and 4-body bound states \cite{moroz2012}. 
		To obtain the RG flow in theory space, we have plotted $g_4$ as a function of $g_3$ in Fig.~\ref{fig:fig2}{\bf b}, by eliminating the explicit cutoff $k$ dependence from $g_3$ and $g_4$. A topological feature of the limit cycle emerges if we regard the 2-dimensional space of $g_3$ and $g_4$ as a torus by gluing the opposite sides of the edges in Fig.~\ref{fig:fig2}{\bf b}: The brown RG limit cycle forms a closed loop, which winds the torus twice in the $g_4$ direction and once in the $g_3$ direction as schematically illustrated in Fig.~\ref{fig:fig2}{\bf c}. These results suggest that the universal numbers of the 3- and 4-body bound states are, in fact, the topological winding numbers of the RG limit cycle. Our finding thus provides the first example of a topological nature of the RG flow in universal few-body physics. An intriguing extension of the present results to the possible topological phase transition in quantum few-body physics will be discussed later.\par
		\subsection{Effective field theory of few-body physics}
		We now present our theoretical framework for obtaining the above results. To make semi-analytic calculations of the 4-body sector possible, we use an effective field theory that exactly reproduces 2- and 3-body observables of identical bosons with the contact ($\delta$-function) interaction. For this purpose, we consider the following action:
		\begin{widetext}
		\begin{eqnarray}
			S[\psi,\phi]&:=&\int_P\psi^*(P)(ip^0+{\bf p}^2-\mu_1)\psi(P)+\int_P\phi^*(P)\left[-\frac{1}{16\pi}\sqrt{\frac{ip^0}{2}+\frac{{\bf p}^2}{4}-\mu_1}-\mu_2\right]\phi(P) \nonumber \\
			&&-\int_{PP_2P_1}G_{\psi}\left(\frac{P}{3}+P_2+P_1\right)\phi^*\left(\frac{2P}{3}+P_2\right)\psi^*\left(\frac{P}{3}-P_2\right)\psi\left(\frac{P}{3}-P_1\right)\phi\left(\frac{2P}{3}+P_1\right),\label{eq:micro-action}
		\end{eqnarray}
		\end{widetext}
		where $P$ denotes the four momentum consisting of Matsubara frequency $p^0$ and momentum ${\bf p}$, $\int_P:=\int\frac{d^4 p}{(2\pi)^4}$, and $G_{\psi}(P):=(ip^0+{\bf p}^2-\mu_1)^{-1}$. In equation~(\ref{eq:micro-action}), $\psi$ describes a particle, and $\phi$ describes a dimer. Throughout this paper, we employ the units $\hbar=2m=1$, where $m$ ($2m$) is the mass of a particle (a dimer). 
		In our model, we have reduced the Yukawa coupling between a particle and a dimer in the ordinary two-channel model of identical bosons \cite{bedaque1999, braaten2006} to the particle exchange interaction, as in the third term on the right-hand side of equation~(\ref{eq:micro-action}). As we show below, our model exactly reproduces the dimer propagator and the Skorniakov-Ter-Martirosian equation \cite{skornyakov1956} for the 3-body scattering. \par
		\subsection{Topologically connected 3- and 4-body limit cycle}
		We employ the functional renormalization-group (FRG) analysis, which allows a non-perturbative RG transformation onto the flowing action $\Gamma_k[\psi,\psi^*,\phi,\phi^*]$, in which the coefficients $\Gamma_{k}^{(3)}$ and $\Gamma_{k}^{(4)}$ of $|\psi\phi|^2$ and $|\psi\psi\phi|^2$ are identified with the $k$-dependent particle-dimer and particle-particle-dimer correlation functions. The exact RG flow of $\Gamma_k[\psi,\psi^*,\phi,\phi^*]$ is governed by the Wetterich equation \cite{wetterich1993} (equation~(\ref{eq:wett}) in Methods) and the flows of $\Gamma_{k}^{(3)}$ and $\Gamma_{k}^{(4)}$ can be extracted from the vertex expansion \cite{morris1994b} (equation~(\ref{eq:vertex-expansion}) in Methods) of the Wetterich equation.\par
		Concerning the 3-body sector, we can exactly solve the FRG equation of $\Gamma_{k}^{(3)}$ (see Fig.~\ref{fig:fig3} in Methods). By focusing on the spatially isotropic $s$-wave component of the scattering amplitude, which makes the dominant contribution in the low-energy Efimov physics, we perform a projection onto $\Gamma_k^{(3)}(p_2,p_1)=2\pi\int d\cos\theta_{p_2p_1}\Gamma_k^{(3)}(ip^0=k^2+3\mu_1;{\bf p}_2,{\bf p}_1)$, and define the dimensionless 3-body coupling constant by
		\begin{eqnarray}
			g_3:=-k^2\Gamma_k^{(3)}(p_2=0,p_1=0).
		\end{eqnarray}
		As in Fig.~\ref{fig:fig2}{\bf a}, we check that $g_3$ exhibits the log-periodic cutoff $k$ dependence, where one period $\Delta\ln k \simeq \ln22.7$ finds good agreement with Efimov's scaling parameter $\pi/s_0\simeq\ln22.6966$, showing that our effective field theory accurately reproduces the self-similarity of the Efimov physics. \par
		Concerning the 4-body sector, we cannot perform an exact FRG calculation of the 4-body sector, since in the total 4-body scattering process, a complicated sub-amplitude of particle-dimer scattering imposes a severe restriction on the numerical calculation; however, we find that the FRG equation of the 4-body sector becomes tractable if we replace the particle-dimer scattering process by the propagation process of a relevant Efimov trimer (see Fig.~\ref{fig:fig4} in Methods). By focusing on the dominant $s$-wave component of the particle-trimer scattering amplitude $T_k^{(4)}(p_2,p_1)=2\pi\int d\cos\theta_{p_2p_1}T_k^{(4)}(ip^0=e^{2\pi/s_0}k^2+4\mu_1;p_2,p_1)$, we define the dimensionless 4-body coupling constant by
		\begin{eqnarray}
			g_4:=\sqrt{E_{T,k}}\;T_k^{(4)}(p_2=0,p_1=0).
		\end{eqnarray}
		Like the 3-body coupling constant $g_3$, the 4-body coupling constant $g_4$ exhibits the log-periodic $k$ dependence with one period of $\Delta\ln k \simeq \ln22.7$. We find that as $g_3$ flows from $-\infty$ to $\infty$, $g_4$ does twice from $-\infty$ to $\infty$, reflecting that there appears two 4-body bound states associated with one Efimov trimer. Indeed, by evaluating the energies of an Efimov trimer and the associated two tetramers from the values of the cutoff $k=k_{3b}$ and $k=k_{4b}^{(1)},k_{4b}^{(2)}$ at which $g_3$ and $g_4$ diverge, respectively, we find the following relations:
		\begin{eqnarray}
			\frac{k_{4b}^{(1)}}{k_{3b}}=1.11,\frac{k_{4b}^{(2)}}{k_{3b}}=3.66,
		\end{eqnarray}
		which agree reasonably with the exact ratio between the energy of an Efimov trimer and that of associated two tetramers $\sqrt{\frac{E_{4b}^{(1)}}{E_{3b}}}=1.00113$, $\sqrt{\frac{E_{4b}^{(2)}}{E_{3b}}}=2.14714$ \cite{deltuva2013a}.\par
		To obtain the RG flow in theory space, we plot $g_4$ as a function of $g_3$ in Fig. 2{\bf b}, by eliminating the explicit cutoff $k$ dependence from $g_3$ and $g_4$. As we can see in Fig. 2{\bf b}, if we regard the two-dimensional space of $g_3$ and $g_4$ as a torus by gluing the edges, the RG limit cycle forms a single closed loop, which winds the torus twice in the $g_4$ direction and once in the $g_3$ direction, clearly indicating that the number of universal clusters corresponds to the topological winding numbers in theory space.
		 
		\subsection{Conclusion \& outlook}
		In conclusion, we have demonstrated that the numbers of universal few-body clusters are, in fact, the topological winding numbers of the renormalization-group flow trajectory in theory space. The one-to-two correspondence in the numbers of Efimov trimers and tetramers is a consequence of the torus topology in the space of the three-body and four-body coupling constants, where the trajectory winds twice in the $g_4$ direction every time it winds in the $g_3$ direction. \par
		An intriguing question is a topological phase transition in the 4-body physics. If we introduce a mass imbalance in a 4-particle system, there is a situation in which the number of 4-body companions accompanying an Efimov trimer changes according to the mass ratio \cite{blume2014}. Our present analysis suggests that such a situation may be regarded as a new type of topological phase transition in theory space at which the winding number of the renormalization-group (RG) limit cycle changes. Now that Efimov trimers are observed in mass-imbalanced mixtures of certain atomic species \cite{barontini2009, pires2014, tung2014}, the topological phase transition in the 4-body physics should be within experimentally reach. \par
		Another important question that arises from the present result is the predicted instability of the tetramers accompanying an Efimov trimer. As pointed out in Ref.~\cite{deltuva2011}, a tetramer can, in principle, decay into a composite of a particle and a deep Efimov trimer, making the tetramer unstable. One may think that such instability rounds off the diverging behavior of the 4-body coupling constant $g_4$ by introducing an imaginary part to $g_4$, and thereby removes the topological feature of the RG limit cycle. However, we find that even if we take the instability into account by introducing a deep Efimov trimer so that a tetramer can decay into the trimer state, the imaginary part of $g_4$ remains zero and the divergence of $g_4$ survives. In addition, if an exact analysis removes the divergence of $g_4$, we expect that we can still define a topological number by using the coupling constant $g_4$ extended to a complex value. Since in the 3-body problem, the limit cycle of complex 3-body coupling constant $g_3$ winds around a fixed point of $g_3$ even when we introduce an artificial imaginary part to $g_3$ \cite{moroz2009}, we expect that the limit cycle of $g_4$ winds around a `node' defined as a parameter-space on which the beta function of $g_4$ vanishes.\par
		\subsection{Methods}
		\subsubsection{Wetterich equation}
		The functional renormalization group (FRG) is described by the Wetterich equation \cite{wetterich1993}:
		\begin{eqnarray}
			\partial_k\Gamma_k[\Phi]=\frac{1}{2}{\rm Tr}\tilde{\partial}_k{\rm ln}\left(\frac{\delta^2\Gamma_k[\Phi]}{\delta\Phi(P)\delta\Phi(P)}+R_{\Phi,k}(P)\right),\label{eq:wett}
		\end{eqnarray}
		where $\Phi(P)=(\psi(P),\psi^*(P),\phi(P),\phi^*(P))$, and $\Gamma_k[\Phi]$ is the flowing action, which is defined as the one-particle irreducible effective action of the cutoff $k$ dependent action $S_k:=S+\int_P\psi^*(P)R_{\psi,k}(P)\psi(P)+\int_P\phi^*(P)R_{\phi,k}(P)\phi(P)$ and reduces in the high-energy limit $k\rightarrow\Lambda$ to the action $S$ and in the low-energy limit $k\rightarrow 0$ to the quantum effective action $\Gamma$ that possesses the full information of Green's functions. The symbol Tr implies the sum over momenta, Matsubara frequency, and field species. The derivative $\tilde{\partial}_k$ acts only on the regulators $R_{\Phi,k}=R_{\psi,k}, R_{\phi,k}$, which are chosen as
		\begin{eqnarray}
			R_{\psi,k}(Q)=\frac{k^2}{c^2},R_{\phi,k}(Q)=\frac{\sqrt{k^2-{\bf q}^2}}{16\pi}\Theta(k^2-{\bf q}^2),\label{eq:reg}
		\end{eqnarray}
		where $\Theta$ is the unit-step function and $c$ is a constant to be specified later. The continuous regulators facilitate numerical calculations. 
		\subsubsection{Vertex expansion}
		To deal with the RG flow of the 3- and 4-body coupling constants, we perform a vertex expansion \cite{morris1994b} of equation~(\ref{eq:wett}) with respect to the field variables to derive the RG equations for one-particle irreducible vertices: 
		\begin{widetext}
		\begin{eqnarray}
			\Gamma_k[\psi,\psi^*,\phi,\phi^*]&:=&\int_P\psi^*(P)G_{\psi,k}^{-1}(P)\psi(P)+\int_P\phi^*(P)\Gamma_k^{(2)}(P)\phi(P) \nonumber \\
			&&+\int_{\substack{P_1,P_2\\P'_1,P'_2}}(2\pi)^4\delta^{(4)}(P_1+P_2-P'_2-P'_1)\Gamma_k^{(3)}(P_1P_2;P'_2P'_1)\phi^*\left(P_1\right)\psi^*\left(P_2\right)\psi\left(P'_2\right)\phi\left(P'_1\right)\nonumber \\
			&&+\frac{1}{(2!)^2}\int_{\substack{P_1,P_2,P_3\\P'_1,P'_2,P'_3}}(2\pi)^4\delta^{(4)}(P_1+P_2+P_3-P'_3-P'_2-P'_1)\Gamma_k^{(4)}(P_1P_2P_3;P'_3P'_2P'_1)\nonumber \\
			&&\hspace{80pt}\times\phi^*\left(P_1\right)\psi^*\left(P_2\right)\psi^*\left(P_3\right)\psi\left(P'_3\right)\psi\left(P'_2\right)\phi\left(P'_1\right)\nonumber \\
			&&+\cdots, \label{eq:vertex-expansion}
		\end{eqnarray}
		\end{widetext}
		where $\Gamma_k^{(n)}$ is the one-particle irreducible vertex that represents the correlation of $n$ particles at the RG cutoff scale of $k$. By substituting equation~(\ref{eq:vertex-expansion}) into equation~(\ref{eq:wett}), we obtain the exact FRG equation of $\Gamma_{k}^{(n)}$. Since we are interested in the 3-body and 4-body physics, we have only to consider the one-particle irreducible vertices up to $n=4$. Indeed, the exact FRG equation of $\Gamma_k^{(n)}$ is closed up to $n$ \cite{floechinger2014}, showing that the $n$-body physics in the vacuum (inverse temperature $\beta=\infty$ and number density $n=0$) is not affected by the $(n+1)$-body physics. \par
		\subsubsection{3-body sector}
		\begin{figure}[b]
  			\begin{center}
  			 	\includegraphics[width=245pt]{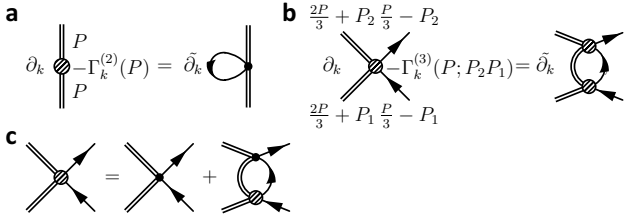}
				\caption{Diagrammatic expressions for the exact FRG equations for {\bf a}, the 2-body sector and {\bf b}, the 3-body sector. The black dot in {\bf a} shows the bare particle exchange interaction introduced in equation~(\ref{eq:micro-action}), the solid curve shows the propagator $G_{\psi}$ of a particle, and the double line shows the propagator $G_{\phi,k}$ of a dimer introduced in equation~(\ref{eq:dprop}). {\bf c}, Integral form of the exact FRG equation for the 3-body sector.}\label{fig:fig3}
  			\end{center}
		\end{figure}
		We first consider the 1-, 2- and 3-body sectors separetely to obtain the limit cycle of the 3-body coupling constant $g_3$. We find that the beta function of the 1-body sector vanishes \cite{braaten2006, moroz2012, horinouchi2015}, because the self-energy correction is absent in the particle vacuum. Therefore, the cutoff $k$ dependent 1-body inverse propagator is given by
		\begin{eqnarray}
			G_{\psi,k}^{-1}(P)=G_{\psi}^{-1}(P)=ip^0+{\bf p}^2-\mu_1.
		\end{eqnarray}
		Concerning the 2-body sector, the exact FRG equation corresponding to Fig.~\ref{fig:fig3}{\bf a} can also be solved analytically. Together with the renormalization condition $\mu_2+\frac{\Lambda}{8\sqrt{2}\pi c}=\frac{1}{16\pi a}=0$ at the unitarity limit $a=\pm\infty$, the exact 2-body sector is given by
		\begin{eqnarray}
			\Gamma_k^{(2)}(P)=\frac{1}{16\pi}\sqrt{\frac{ip^0}{2}+\frac{{\bf p}^2}{4}-\mu_1},\label{eq:2b1pi}
		\end{eqnarray}
		where we follow the trick of Ref.~\cite{diehl2008} to set the constant $c$ in equation~(\ref{eq:reg}) to be $c=\infty$, thereby integrating out the $\psi$-field first and then the $\phi$-field in the RG flow. We note that equation~(\ref{eq:2b1pi}) is consistent with the expression of the dimer self-energy \cite{braaten2006, moroz2009}. The resulting inverse dimer propagator thus becomes
		\begin{eqnarray}
			G_{\phi,k}^{-1}(P)=R_{\phi,k}(P)+\frac{1}{16\pi}\sqrt{\frac{ip^0}{2}+\frac{{\bf p}^2}{4}-\mu_1}.\label{eq:dprop}
		\end{eqnarray}
		The 3-body sector can also be solved exactly. The exact FRG equation for the 3-body sector corresponding to Fig.~\ref{fig:fig3}{\bf b} can be analytically integrated with respect to the cutoff $k$, resulting in an integral equation corresponding to Fig.~\ref{fig:fig3}{\bf c}. We note that the integral equation reduces in the infrared limit $k\rightarrow 0$ to the Skorniakov-Ter-Martirosian equation \cite{skornyakov1956}. Since the $s$-wave component of the scattering amplitude makes the dominant contribution to the low energy Efimov physics, we perform a projection onto $\Gamma_k^{(3)}(p_2,p_1)=2\pi\int d\cos\theta_{p_2p_1}\Gamma_k^{(3)}(ip^0=k^2+3\mu_1;{\bf p}_2,{\bf p}_1)$, and define the dimensionless 3-body coupling constant by $g_3:=-k^2\Gamma_k^{(3)}(p_2=0,p_1=0)$.By solving the FRG equation corresponding to Fig.~\ref{fig:fig3}{\bf c}, we obtain the exact RG flow of $g_3$.\par
		\subsubsection{4-body sector}
		\begin{figure}[t]
  			\begin{center}
  			 	\includegraphics[width=240pt]{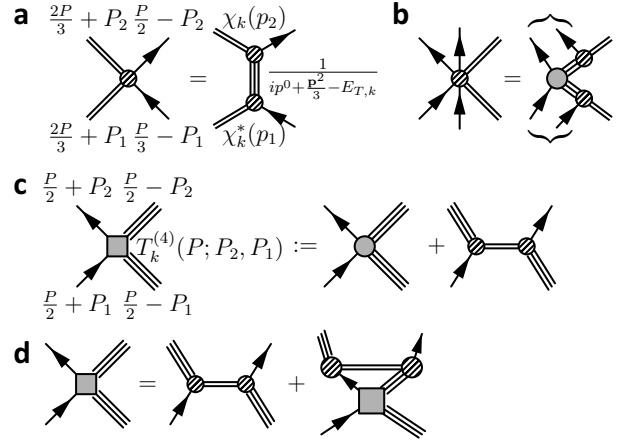}
				\caption{Decomposition of {\bf a}, the 3-body one-particle irreducible vertex and {\bf b}, the 4-body one-particle irreducible vertex. The curly brackets indicate symmetrization with respect to external lines. {\bf c}, Definition of the particle-trimer scattering amplitude $T_k^{(4)}$ represented as a square vertex. (d) Integral form of the exact FRG equation for the 4-body sector.}\label{fig:fig4}
  			\end{center}
		\end{figure}
		Concerning the 4-body sector, we cannot perform an exact FRG calculation since the 4-body sector requires the knowledge of the full momentum dependence of the particle-dimer scattering amplitude, which is too complicated to deal with directly. Our strategy is to take into account the dominant intermediate state in the total particle-dimer scattering process by making use of the K\"all\'en-Lehmann spectral representation: By focusing on the $s$-wave component of the particle-dimer scattering amplitude, we decompose $-\Gamma_k^{(3)}(ip^0,{\bf p};p_2,p_1)$ as
		\begin{eqnarray}
			-\Gamma_k^{(3)}(ip^0,{\bf p};p_2p_1)=\frac{\chi_k(p_2)\chi_k^*(p_1)}{ip^0+\frac{{\bf p}^2}{3}-E_{T,k}},\label{eq:sep-3b}
		\end{eqnarray}
		where $E_{T,k}$ is the binding energy of an intermediate Efimov state and $\chi_k(p)$ is the Bethe-Salpeter (bound-state) wave function of the Efimov trimer. Diagrammatically, equation~(\ref{eq:sep-3b}) can be expressed as shown in Fig.~\ref{fig:fig4}{\bf a}. Equation~(\ref{eq:sep-3b}) means that we replace the particle-dimer scattering process by a propagation process of a relevant Efimov state. This approximation is based on the same idea as the separable pole approximation \cite{alt1970, grassberger1967}, which enables us to deal with the particle-dimer scattering amplitude with a factorized momentum dependence as in equation~(\ref{eq:sep-3b}), and has played a crucial role in the accurate computation of the quantum few-body problems.\par
		The function $\chi_k$ and $E_{T,k}$ can be determined from the Bethe-Salpeter (bound-state) equation which is obtained by substituting equation~(\ref{eq:sep-3b}) into the 3-body equation as depicted in Fig.~\ref{fig:fig3}{\bf c}. If we choose the intermediate Efimov state such that $E_{T,k}\gg k^2$, the Bethe-Salpeter equation reduces to an analytically solvable one given in Ref.~\cite{gogolin2008}, and the resulting $\chi_k$ and $E_{T,k}$ are given by
		\begin{eqnarray}
		&&\chi_k(p)=A\frac{{\rm sin}\left[s_0{\rm arcsinh}\left(\frac{\sqrt{3}p}{\sqrt{2E_{T,k}}}\right)\right]}{p/\sqrt{E_{T,k}}},\label{eq:chi}\\
		&&E_{T,k}=6\Lambda^2e^{-\frac{2n(k)\pi}{s_0}},\label{eq:etk}
		\end{eqnarray}
		where $s_0\simeq 1.00624$ is Efimov's scaling parameter, $n(k)$ is an integer, $\Lambda$ is the ultraviolet cutoff, and $A\simeq 5.00858$ is the normalization factor determined through the procedure developed in Ref.~\cite{lurie1965}. 
		To take into account the relevant intermediate Efimov state that satisfies $E_{T,k}\gg k^2$, we choose the integer $n(k)$ in equation~(\ref{eq:etk}) as 
		\begin{eqnarray}
			n(k)=\left\lfloor\frac{s_0}{2\pi}{\rm log}\frac{6\Lambda^2}{k^2}\right\rfloor, \label{eq:cn}
		\end{eqnarray}
		where $\lfloor x\rfloor$ refers to the floor function which gives the largest integer less than $x$.. \par
		Based on equations~(\ref{eq:sep-3b})-(\ref{eq:cn}), the 4-body sector is greatly simplified. Following Ref.~\cite{tanizaki2013}, we decompose the 4-body one-particle irreducible vertex as depicted in Fig.~\ref{fig:fig4}{\bf b}. The resulting 4-body FRG equation can be integrated with respect to $k$, resulting in a simple form as depicted in Fig.~\ref{fig:fig4}(d). We note that the integrated 4-body FRG equation possesses the same structure as the Alt-Grassberger-Sandhas equation \cite{grassberger1967, alt1970}, except that our equation lacks the dimer-dimer scattering process, which is irrelevant at the unitarity limit where the dimer state is absent. As in the 3-body sector, we perform an $s$-wave projection onto $T_k^{(4)}(p_2,p_1)=2\pi\int d\cos\theta_{p_2p_1}T_k^{(4)}(ip^0=e^{2\pi/s_0}k^2+4\mu_1;p_2,p_1)$ and define the dimensionless 4-body coupling constant by $g_4:=\sqrt{E_{T,k}}T_k^{(4)}(p_2=0,p_1=0)$.	 By solving the FRG equation given in Fig.~\ref{fig:fig4}(d), we obtain the non-perturbative RG flow of $g_4$, and thus we obtain Fig.~\ref{fig:fig2}.\par
		\subsection{Acknowlegement}
		This work was supported by KAKENHI Grant No. 26287088 from the Japan Society for the Promotion of Science (JSPS), a Grant-in-Aid for Scientific Research on Innovative Areas ``Topological Materials Science" (KAKENHI Grant No. 15H05855), the Photon Frontier Network Program from MEXT of Japan, and the Mitsubishi Foundation. Y.H. is supported by a Grants-in-Aid for JSPS Fellows (No. 15J11136).  All authors equally contributed to this work and declare no competing financial interests. Correspondence and requests for materials should be addressed to Y. H.

\end{document}